\documentclass[prl,twocolumn,preprintnumbers,amsmath,amssymb]{revtex4}

\usepackage{graphicx}

\def\be{\begin{equation}}
\def\ee{\end{equation}}
\def\l{\lambda}
\def\cb{\bar{c}}
\def\d{\delta}
\def\o{\omega}

\def\O{{\cal O}}
\def\bea{\begin{eqnarray}}
\def\eea{\end{eqnarray}}

\begin{document}

\title{Correlation functions of the one-dimensional attractive Bose gas}
\author{Pasquale Calabrese${}^{1}$ and Jean-S\'ebastien Caux${}^{2}$}
\affiliation{$^{1}$Dipartimento di Fisica dell'Universit\`a di Pisa and INFN, 
Pisa, Italy}
\affiliation{$^2$Institute for Theoretical Physics, Universiteit van Amsterdam,
1018 XE Amsterdam, The Netherlands}

\date{\today}

\begin{abstract}
The zero-temperature correlation functions of the 
one-dimensional attractive Bose gas with delta-function interaction are
calculated analytically for any value of the interaction parameter and number of particles,
directly from the integrability of the model.
We point out a number of interesting features, including zero recoil energy for large
number of particles, analogous to a M\"ossbauer effect.  
\end{abstract}

\maketitle

%{\it Introduction}.
Recent experiments on trapped one-dimensional (1D) atomic gases
\cite{exprep,expattr} provide a unique opportunity to study the effects of
quantum correlations and fluctuations for one of the 
paradigms of strongly-correlated systems:  the Lieb-Liniger interacting Bose 
gas \cite{LL}. One of the striking features of these experiments is that 
the effective 1D coupling $c$ can be tuned to essentially any (positive or 
negative) value.  
On the theoretical side the model is exactly solvable, and therefore physics
beyond mean-field can be reliably investigated.  Although the repulsive regime 
is well-understood, the attractive case is less commonly discussed in the literature.
This case is unusual because the atoms can bind together:  in fact, the
ground state of $N$ particles of mass $m$ in infinite space is a clump
whose binding energy is $E_0=-m c^2N(N^2-1)/6\hbar^2$ \cite{mg-64}.
This clump behaves like a particle of mass given by $Nm$.

Our purpose here is to provide a detailed analysis of the dynamics of this
system, by calculating observable correlation functions analytically.  
We give expressions for the
one-body function and dynamical structure factor (DSF), which
are experimentally accessible using ballistic 
expansion and Bragg spectroscopy, and highlight their 
important features.  
The attractive regime has found renewed experimental interest with the observation of bright solitons 
in quasi 1D harmonically trapped condensates \cite{expattr}.

The Hamiltonian of the Lieb-Liniger model is given by
\bea
H = -\frac{\hbar^2}{2m}\sum_{j=1}^N \frac{\partial^2}{\partial {x_j^2}} 
- 2\cb \sum_{\langle i,j \rangle} \delta(x_i - x_j)
\label{LL}
\eea
with $\cb=-c > 0$ the interaction parameter, 
$m$ the mass of the particles (atoms), and the sum runs over all pairings.
In terms of experimental parameters the 1D coupling constant is 
$\cb=\hbar^2/m a_{1D}$, where $a_{1D}$ is the effective 1D 
scattering length that can be tuned via Feshbach resonance or
transverse confinement \cite{o-98}.
For definiteness, we consider a system of length $L$ with periodic 
boundary conditions. 
From now on we fix $\hbar=2m=1$.

Hamiltonian (\ref{LL}) is diagonalized by the Bethe Ansatz \cite{LL}: 
the eigenfunctions are superpositions of plane waves, 
$\Psi = \sum_P A_P \prod_{i=1}^N e^{i \l_{P_j} x_j}$ over all permutations $P$ of the momenta 
(rapidities) $\l$.  The $A_P$ coefficients are functions of the two-particle scattering 
phase shifts obtained from (\ref{LL}), and periodicity of the wavefunction requires
the set of rapidities $\{ \l \}$ to be solution to the Bethe equations
\be
e^{i\l_a L}=\prod_{a\neq b} \frac{\l_a-\l_b-i\cb}{\l_a-\l_b+i\cb}\,, ~~~a = 1,...,N.
\label{BE}
\ee
For the repulsive case, all solutions are real \cite{yy}, and the wavefunctions are scattering
states of $N$ atoms.  For the attractive case $\cb > 0$, however,
the physics is completely different.  Complex solutions to
the Bethe equations are allowed:  atoms can therefore bind
and form new types of particles which are stable under scattering.
A general eigenstate is therefore made by partitioning the $N$ atoms into a set of $N_s$ 
bound-states.  The rapidities associated to a bound-state of $j$ atoms
form a regular pattern in the complex plane which is called a {\it string} \cite{Tbook}:
\be
\l^{j, a}_{\alpha}=\l^j_{\alpha} +\frac{i\cb}2(j+1-2a)+i\d^{j,a}_{\alpha}\,.
\label{strrap}
\ee
Here, $a = 1,...,j$ labels the rapidities within the string, while $\alpha = 1, ..., N_j$
labels the set of $N_j$ strings of length $j$.  Note that $N = \sum_{j} j N_j$, with the total 
number of strings given by $N_s = \sum_j N_j$.  $\d^{j,a}_{\alpha}$ are
deviations which fall off exponentially with system size $L$.
Perfect strings (i.e. with $\d=0$) are then
exact eigenstates in the limit $L\to\infty$ for arbitrary $N$.
Thus we consider the limit $L\to\infty$ at fixed $N$
\footnote{This limit is not trivial (as it is in the repulsive regime):  here, the $N$ particles
remain strongly correlated and bound to one another even when $L \to \infty$.}.

Such an eigenstate will have momentum and energy
\be
K=\sum_{(j,\alpha)} j \lambda^j_{\alpha} \,,\quad
E = \sum_{(j,\alpha)} \left[j {\lambda^j_{\alpha}}^2 - \frac{\cb^2}{12} j(j^2 - 1)\right],
%E=\sum_j\left[n_j k_j^2-\frac{\cb^2}{12} n_j(n_j^2-1)\right]\,.
\label{strE}
\ee
where $\sum_{(j,\alpha)}$ represents the sum over strings in the eigenstate.
The set of $N_s$ rapidities $\lambda^j_{\alpha}$ obey a set of reduced
Bethe equations obtained by substituting (\ref{strrap}) into (\ref{BE}).

The ground-state corresponds to a single $N$-string centered on zero.
Excited states can be classified in terms of their string
content. 
{\it One-particle} (or one string $N_s=1$) states are obtained by giving a 
finite momentum $K$ to the ground state. Their dispersion relation is $\o=K^2/N$,
meaning that in the large $N$ limit they form a flat dispersion band degenerate with the ground state.
{\it Two-particle} states ($N_s=2$) are composed 
of two strings made of $M$ and $N-M$ particles respectively.
{\it Multiparticle} states with $N_s>2$ are similarly constructed.

{\it Form factors and correlation functions}.
We consider the zero-temperature density-density correlation function 
$S^\rho(x,t)=\langle\rho(x,t)\rho(0,0)\rangle$ with 
$\rho(x,t) =\sum_{j=1}^N \d(x-x_j)$, whose Fourier transform is known as the dynamical 
structure factor (DSF), and the 
one-body dynamical correlation function of the canonical Bose field $\Psi(x,t)$,
$S^\Psi(x,t)=\langle\Psi^\dag(x,t) \Psi(0,0)\rangle$.
By Fourier transform ${\cal O} (k,\omega) = \int_0^L dx \int_{-\infty}^{\infty} dt e^{i(\omega t - kx)} {\cal O}(x,t)$,
these can be written as a sum over intermediate states $\mu$, 
\bea
S^\O(k,\o)= 2\pi L \sum_\mu \frac{|\Sigma^{\O}_\mu|^2}{||GS||^2 ||\mu||^2}
\d(\o-E_\mu+E_0)\,,
\label{Lehmann}
\eea
where $GS$ denotes the ground state, and the form factor (FF) $\Sigma^{\O}_\mu =\langle \mu|\O(0,0)|GS\rangle
= \frac{1}{L} \langle \mu | \O_{K_{\mu}} | GS \rangle$ 
depends on the operator $\O$ and on the state $\mu$
($\O_K$ is the Fourier transform of the operator $\O$ at momentum $K$).
$|| \mu ||$ denotes the norm of state $\mu$. 
For the Lieb-Liniger model, FFs and norms were calculated for 
general Bethe states in Refs. \cite{KBOOK,kks-97,s-89,cc-06g}. 
They are given by the determinant of matrices whose entries are 
rational functions of the rapidities of the two eigenstates involved 
(we do not report these rather complicated expressions here).

In the repulsive case, the Bethe equations cannot be solved analytically.
It is however possible to solve the Bethe equations numerically at finite $N$ and $L$, 
determine the corresponding FFs, and perform the sum over intermediate states 
by selectively scanning the Fock space, thereby obtaining very precise results
for the correlation functions \cite{cc-06,cc-06g}.
In the present case, however, the rapidities can be determined to
exponential precision for all states such that $N_s\ll N$.  The remaining
$N_s$ equations for the string centers yield rapidities which in
the limit $L\to\infty$ are quantized as for free particles (full details will be published 
elsewhere \cite{prep}).  This allows us to go analytically much further than in
the repulsive case.

One technical difficulty is that 
the elements of the norm and FF matrices are singular when 
calculated on exact strings with $\d=0$.  A more careful analysis
however shows that all divergences cancel \cite{prep}, and that
it is possible to obtain closed forms for all norms and form factors.  
Single-particle and density FFs between the 
ground state and any other string state have the rather simple forms
\bea\displaystyle
|\Sigma_\mu^\Psi|&=& \cb^{\frac{1}{2}} N!(N-1)! \displaystyle\prod_{(j,\alpha)}
H_{j} (\mu^j_{\alpha}/\bar{c}), 
\label{FFg}\\
|\Sigma_\mu^\rho|&=& \frac{K_{\mu}^2}\cb N!(N-1)! \displaystyle\prod_{(j,\alpha)}
H_{j} (\mu^j_{\alpha}/\bar{c}), 
\label{FFs}
\eea
where we have defined the fundamental block $H_{M} (x) = \left|\frac{\Gamma(\frac{N-M}{2} + ix)}
{\Gamma(\frac{N+M}{2} + ix)}\right|^2$.
These expressions are exact in the limit of large $L$ for any finite $N$ \cite{foot2}.
The general expression for the norms is
\be
||\mu||^2 = (L\cb)^{N_s} \prod_j j^{2N_j} \prod_{(j,\alpha) > (k,\beta)} F^{jk}(\mu_{\alpha}^j - \mu_{\beta}^k)
\label{norms}
\ee
where $F^{jk} (\mu) = (\frac{\mu^2}{\cb^2} + \frac{(j+k)^2}{4})/(\frac{\mu^2}{\cb^2} + \frac{(j-k)^2}{4})$.

Eqs. (\ref{FFg}), (\ref{FFs}) and (\ref{norms}) give an 
exact representation of each term in the expansion (\ref{Lehmann}) of 
the desired correlation functions for $L\to\infty$ for 
any value of $N$ and $\cb$. 
The sums over intermediate states can now be analytically performed for various families 
of excited states, starting from the simplest ones.

{\it Dynamical structure factor}.
Let us start here with the simplest excited states, which are composed of one $N$-string.
For arbitrary $N$, these states give a single coherent peak,
\be
S^{\rho}_{0} (k,\o)= 2\pi \frac{N^2}{L} \d(\o-\frac{k^2}N)
\prod_{a=1}^{N-1}[1 + k^2/a^2\cb^2N^2]^{-2}.
\label{S_rho_N_gen}
\ee
The leading multiparticle contributions come from two-particle $N-M:M$ states, for which we find
\be
S^{\rho}_M (k, \omega) \!=\! \frac{\Theta(\omega \!-\! \omega^{\rho}_{M}) k^4 \Gamma^4(N)}
{[\omega \!-\! \omega^{\rho}_{M}]^{\frac{1}{2}}\bar{c}^5 C^N_M} 
\!\!\sum_{\sigma = \pm}\!\! \frac{H^2_{N\!-\!M}(\frac{\mu^{\sigma}_s}{\bar{c}}) H^2_{M} (\frac{\mu^{\sigma}_M}{\bar{c}})}
{F^{N-M,M}(\mu^{\sigma}_s \!-\! \mu^{\sigma}_M)}
\ee
where $\Theta(x)$ is the Heaviside step function,
$C^N_M = 2 L N^{\frac{1}{2}} [(N-M)M]^{\frac{3}{2}}$ and 
$\omega^{\rho}_{M}(k) = \frac{\cb^2}{4} NM(N-M) + \frac{k^2}{N}$ is the lower threshold
for these excitations (there is no upper threshold).
$\mu_s$ and $\mu_M$ are respectively the rapidities of the $N-M$ and $M$ strings.
Energy and momentum conservation enforce 
$\mu_s^{\pm} = \frac{k}{N} \mp [\frac{M}{N(N\!-\!M)}]^{\frac{1}{2}}
[\omega - \omega^{\rho}_{M}(k)]^{\frac{1}{2}}$, $\mu_M^{\pm} = \frac{k}{N} \pm [\frac{N\!-\!M}{NM}]^{\frac{1}{2}}
[\omega - \omega^{\rho}_{M}(k)]^{\frac{1}{2}}$. 

{\it One-body Green function}.
The most important intermediate states here 
are those composed of one $N-1$ string with total momentum $k$. 
For arbitrary $N$, the contribution of these states to the one-body function is
\be
S^{\Psi}_{0} (k, \omega) = \frac{2\pi/(\bar{c}L(N-1)^{2}) 
~\delta (\omega - \frac{k^2}{N-1})}{\prod_{a=1}^{N-1} [(1 - \frac{1}{2a})^2 + (\frac{k}{a\bar{c}(N-1)})^2]^2}.
\ee
The leading multiparticle contributions come from the two-particle $N-1-M:M$ states,
\be
S^{\Psi}_{M} (k,\omega) \!=\! \frac{\Theta(\omega \!-\! \omega^{\Psi}_{M}) \Gamma^4(N)}
{[\omega \!-\! \omega^{\Psi}_{M}]^{\frac{1}{2}} \bar{c}^2 C^{N-1}_M} 
\!\!\sum_{\sigma = \pm}\!\! \frac{H^2_{N\!-\!1\!-\!M}(\frac{\mu^{\sigma}_s}{\bar{c}}) H^2_{M} (\frac{\mu^{\sigma}_M}{\bar{c}})}
{F^{N-1-M,M}(\mu^{\sigma}_s \!-\! \mu^{\sigma}_M)}.
\ee
%where $C_M^{\Psi} = \frac{\Gamma^4(N)}{2\bar{c}^2 L (N-1)^{1/2} [(N-1-M)M]^{3/2}}$.
These form a continuum extending above the boundary 
%$\omega^{\Psi}_{M} (k) \!\!=\!\! \frac{k^2}{N\!-\!1} \!+\! \frac{\bar{c}^2}{4} (N\!-\!1)(N\!-\!1\!-\!M)M$,
$\omega^{\Psi}_{M} (k)$ (obtained from $\omega^{\rho}_M$ by replacing $N \rightarrow N-1$).
Energy and momentum conservation constrain $\mu_s, \mu_M$ as for $S^{\rho}$ (replacing $N$ by $N-1$ 
and $\omega^{\rho}_{M}$ by $\omega^{\Psi}_{M}$).
%$\mu_s^{\pm} = \frac{k}{N\!-\!1} \mp [M/((N\!-\!1)(N\!-\!1\!-\!M))]^{1/2}
%[\omega - \omega^{\Psi}_{M,l}(k)]^{1/2}$, $\mu_M^{\pm} = \frac{k}{N\!-\!1} \pm [(N\!-\!1\!-\!M)/(N\!-\!1)M]^{1/2}
%[\omega - \omega^{\Psi}_{M,l}(k)]^{1/2}$.

All other families of excited states can be similarly considered, and the desired correlation
function can be reconstructed to any accuracy required 
\footnote{In general, the contribution from a family of $P$ particle excited states is a
$P-2$-fold integral.  All terms have $1/L$ dependence, which can be traced to our
choice of normalizing the intermediate states to $1$ rather that $L$.}.

We now consider the special large $N$ limit obtained by sending $N\to\infty$ while keeping 
$g=\cb N$ constant.  This prescription maintains the binding energy per particle 
$E_0/N=-g^2/12$ finite \cite{cd-75}.  
%At constant $g$ the norms simplify to
%$||\mu||^2=(\cb L)^{N_s}\prod_{j=1}^{N_s} n_j^2$,
%and the correlation functions simplify because, as we will show, we only need states 
%with at maximum $N_s=2$ for which we can write down very simple analytical results.
In this limit, the correlators mentioned simplify considerably.
For the DSF, we find that the coherent mode contribution becomes
\be
S^{\rho}_0 (k, \omega) = 2\pi \frac{N^2}{L} \frac{(\pi k/g)^2}{\sinh^2 \pi k/g} \delta (\omega - \frac{k^2}{N}).
\label{S_rho_N}
\ee
and for multiparticle states (the leading $M=1$ part),
\bea
S^{\rho}_{1} (k,\omega) = \frac{N}{L} \frac{k^4}{2 g} \frac{\Theta(\omega - g^2/4)}
{[\omega - g^2/4]^{\frac{1}{2}} \omega^2} \sum_{\sigma=\pm} F_1^2 (x_{\sigma})
\label{S_rho_Nm1_1}
\eea
with $F_1(x) = \frac{\pi}{\cosh(\pi x)}$ and $x_{\pm} = ([\omega - g^2/4]^{\frac{1}{2}} \pm k)/g$.

To certify our results, a number of checks can be made.
The first is to compute the second density moment $\langle\rho^2\rangle$,
which is given by the Hellmann-Feynman theorem as $\frac{-1}{L}\frac{\partial E_0}{\partial \bar{c}} 
= \frac{\bar{c}N^3}{6L} + ...$.  Integrating shows that the one-particle part 
$S^{\rho}_0 (k,\omega)$ completely saturates this to leading order in $N$.
To go further, we can study the $f$-sumrule, stating that at 
fixed $k$ the integral  $\int_{-\infty}^{\infty} \frac{d\o}{2\pi} \o S(k,\o)$
equals $k^2 N/L$.  Thus, in the large $N$ limit, the one $N$-string states contribute to the $f$-sumrule as
\be
f_0(k)=\frac{(\pi k/g)^2}{\sinh^2(\pi k/g)}\frac{N}{L} k^2 < \frac{N}{L} k^2\,,
\label{f0}
\ee
and therefore to achieve complete saturation, we here need higher states (this is natural, since
we are computing a first frequency moment).  The leading $M=1$ two-particle states contribute
to the $f$-sumrule as
\be
f_{1}(k)=\frac{N k^2}L  \left(1 - \frac{(\pi k/g)^2}{\sinh^2(\pi k/g)} \right)
\ee
and therefore completely saturate the remaining part of the $f$-sumrule when 
$N$ string contributions have been taken into account.  Higher frequency moments further
suppress the $S^{\rho}_0$ part;  in general, the full DSF is therefore well approximated by
$S^{\rho}(k,\omega) = S^{\rho}_0 (k,\omega) + S^{\rho}_1 (k,\omega)$.

The static structure factor is the integral of the dynamical one and can be 
written as $S^\rho(k)=\int \frac{d\o}{2\pi} S^\rho(k,\o)= S^{\rho}_0(k)+S^{\rho}_{1}(k)$, 
with $S^{\rho}_0 (k)$ trivial and 
\be
S^{\rho}_{1}(k)= \frac{2Nk^4}{Lg^4} \left[\frac{g^2}{k^2}
-\frac{\pi^2}{\sinh^2 \pi \frac{k}{g}}-{\rm Re} ~\psi_2\!\left(i \frac{k}{g}\right)\right]
\ee
where $\psi_2(z)$ is the polygamma function (second 
derivative of the logarithm of the Gamma function). 
For large $k$ the static structure factor is exponentially suppressed and is 
dominated by the two-particle term. The latter shows a peak at $k/g\sim1.2$
reminiscent of what is obtained for the super-Tonks-Girardeau gas-like regime
\cite{abcg-05}.  

For the one-body function, we similarly find 
\be
S^\Psi_{0}(k,\o)=\frac{2\pi^3}{g\cosh^2 (\pi k/g)}\frac{N}{L}\delta(\o-k^2/N)\,
\label{S_Psi_Nm1}
\ee
with two-particle states with $M=1$ giving the leading contribution in the continuum,
\be
S^{\Psi}_1 (k,\omega) = \frac{g^2}{2L} \frac{\Theta(\omega - g^2/4)}
{[\omega - g^2/4]^{\frac{1}{2}} \omega^2} \sum_{\sigma=\pm} F_2^2 (x_{\sigma})
\label{S_Psi_Nm2_1}
\ee
with $F_2(x) = \pi x/\sinh \pi x$ and $x_{\pm}$ as defined above.
To check the relative weight of these states, we use the sumrule that the momentum and
frequency integral of this must equal the density $N/L$, and obtain
$\int_{-\infty}^{\infty} \frac{dk}{2\pi} \int_{-\infty}^{\infty} \frac{d\omega}{2\pi}
S^{\Psi}_{0} (k, \omega) = \frac{N}{L},$
%\be
%\frac{L}{2\pi} \int_{-\infty}^\infty dK \frac{\pi^2}{g\cosh^2 (\pi K/g)}\frac{N}{L^2}
%=\frac{N}L\,,
%\ee
i.e. the one-particle states completely saturate this sumrule.  However, as
we have seen with the DSF, calculating
higher frequency moments would necessitate taking the multiparticle terms into account,
starting with the $M=1$ part.  This correlation function is therefore well-approximated
by $S^{\Psi}(k,\omega) = S^{\Psi}_0(k,\omega) + S^{\Psi}_1 (k,\omega)$.

The sum rules we have studied are saturated using only 
either the one-particle intermediate states, or the simplest
two-particle excitations formed by breaking the original string in
two pieces, one of which is a single rapidity.  For large $N$,
the contributions from higher excitations are suppressed by
progressively higher powers of $1/N$, and can be
neglected for most practical purposes.  The importance of 
the simplest two-particle excitations is highlighted by
calculating higher frequency moments.

{\it Discussion of the results}.
Both correlation functions considered here show similar features.
They are dominated by single-particle coherent contributions,
lying on lines $\omega = k^2/N$ which become dispersionless 
in the large $N$ limit.  There however exist multiparticle continua starting 
from finite mass gap thresholds ($\omega_l \sim g^2/4$ for the leading two-particle ones) and extending to
arbitrarily high energies.  At the lower threshold, the correlators are square-root singular.
At high energies, the correlators
decay exponentially, which is reminiscent of massive
integrable quantum field theories \cite{EsslerREVIEW}.

The fact that the flat band single-particle excitations account for most of the
correlation functions can interestingly be interpreted as a M\"ossbauer-like effect, where the recoil energy
(say upon scattering with a photon in Bragg spectroscopy) vanishes
because the system acts as a single particle of mass $N \gg 1$
(the gound state string can in fact be viewed as a crystal in rapidity space).  It should therefore
be feasible to observe an analogue of the M\"ossbauer effect on this system.

\begin{figure}
\includegraphics[width=8cm]{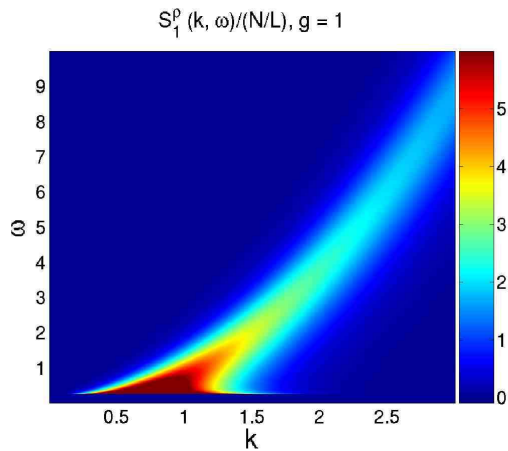} 
\caption{Contribution to the DSF coming from $N-1:1$ (two-particle) states,
in the large $N$ limit for $g = 1$.  The square-root singularity at the lower
threshold is accompanied by a maximum around $k = [\omega - g^2/4]^{\frac{1}{2}}$.}
\label{S1fig}
\end{figure}

\begin{figure}
\includegraphics[width=8cm]{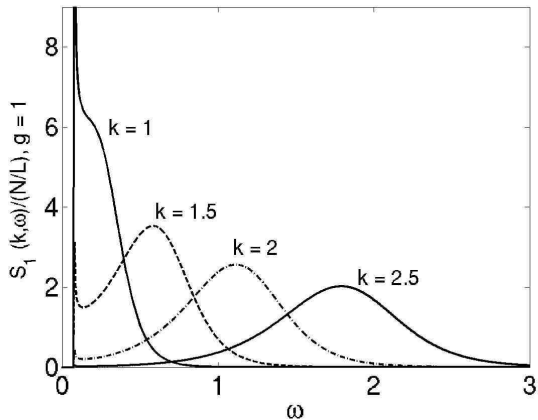} 
\caption{Fixed momentum cuts of Fig. \ref{S1fig}, showing the weakening of
the square-root singularity at higher momentum, and the displacement of the
maximum towards higher energies.}
\label{S1_k_fig}
\end{figure}

The two-particle ($N-1:1$) part of the DSF is plotted in Figure \ref{S1fig} for large $N$,
with $g = 1$.  At the lower threshold, the DSF diverges as $(\omega - g^2/4)^{\frac{1}{2}}$.
For $\o>g^2/4$  the DSF is a monotonous decreasing function of $\o$
as long as $k/g<x_c=1.0565\dots$, 
whereas for $k/g>x_c$ it shows a characteristic broad peak, 
whose position grows like $k^2$ for large $k$ and its amplitude decreases 
like $k^{-1}$.
Away from this peak the correlator decays exponentially.  
Four fixed momentum cuts are
given in Figure \ref{S1_k_fig}, showing these features in more detail.  
For the one-body correlator, the maximum lies at the lower threshold, with monotonous
exponential decay at higher energies.

{\it Concluding remarks}.
In this letter, we showed how integrability could be used to compute the zero-temperature dynamical 
structure factor (Eqs (\ref{S_rho_N}), (\ref{S_rho_Nm1_1})) 
and the dynamical one-body correlation function (Eqs (\ref{S_Psi_Nm1}),(\ref{S_Psi_Nm2_1})) 
for the 1D attractive Bose gas.
The corresponding static correlation functions are obtained as a subset 
of our results.  The results obtained clearly illustrate that 
this experimentally-realizable quantum-mechanical system has many similarities with
massive integrable quantum field theories, namely the correlation functions are given 
(for fixed $k$, as a function of $\omega$) by a coherent peak followed by a multiparticle continuum.
Additionally, we have quantified an interesting zero recoil energy effect for large particle
numbers, which is analogous to the M\"ossbauer effect in solids.

The methods and results obtained in this 
letter can be adapted to describe many other features 
of the attractive Lieb-Liniger gas. 
For example, our expressions could lead 
to a calculation of finite temperature correlation functions, allowing 
the description of the quantum to classical crossover in this strongly interacting model. 

The authors would like to acknowledge interesting discussions with G. Mussardo and
F. H. L. Essler, and support from the Stichting voor Fundamenteel Onderzoek der Materie
(FOM) of the Netherlands.


\begin{thebibliography}{99}
\bibitem{foot2}
Note that, although very similar, the two FFs are very
different. In fact, the operator $\rho$ conserves the particle
numbers whereas $\Psi$ decreases it by 1. So $N_j$ in these equations fulfills
different constraints $\sum_j N_j=N$ and $\sum_j N_j=N-1$ 
for $\rho$ and $\Psi$ respectively.
\end{thebibliography}
\end{document}